\documentclass{aip-cp}

\usepackage[numbers]{natbib}
\usepackage{rotating}
\usepackage{graphicx}
\usepackage{siunitx}
\DeclareSIUnit\electron{e^{-}}
\usepackage{textgreek}

\usepackage{tikz}
\newcommand\copyrighttext{%
  \footnotesize This article may be downloaded for personal use only. Any other use requires prior permission of the author and AIP Publishing. The following article appeared in AIP Conf. Proc. \textbf{2054}, 060062 (2019) and may be found at https://doi.org/10.1063/1.5084693 .}
\newcommand\copyrightnotice{%
\begin{tikzpicture}[remember picture,overlay]
\node[anchor=south] at (current page.south) {\fbox{\parbox{\dimexpr\textwidth-\fboxsep-\fboxrule\relax}{\copyrighttext}}};
\end{tikzpicture}%
}

\makeatletter 
\renewcommand\@biblabel[1]{#1.} 
\makeatother

\begin{document}

% Title portion
\title{Performance of ePix10K, a High Dynamic Range,\\Gain Auto-Ranging Pixel Detector for FELs}

\author[aff1]{G.~Blaj\corref{cor1}}%\noteref{note1,note2}}
\author[aff1]{A.~Dragone}
\author[aff1]{C.~J.~Kenney}
\author[aff1]{F.~Abu-Nimeh}
\author[aff1]{P.~Caragiulo}
\author[aff1]{D.~Doering}
\author[aff1]{M.~Kwiatkowski}
\author[aff1]{B.~Markovic}
\author[aff1]{J.~Pines}
\author[aff1]{M.~Weaver}
\author[aff1]{S.~Boutet}
\author[aff1,aff2]{G.~Carini}
\author[aff1]{C.-E.~Chang}
\author[aff1]{P.~Hart}
\author[aff1]{J.~Hasi}
\author[aff1]{M.~Hayes}
\author[aff1]{R.~Herbst}
\author[aff1]{J.~Koglin}
\author[aff1]{K.~Nakahara}
\author[aff1]{J.~Segal}
\author[aff1]{G.~Haller}

%\eaddress[url]{http://www.aip.org}
%\author[aff2,aff3]{Author's Name\noteref{note2}}
%\eaddress{anotherauthor@thisaddress.yyy}

\affil[aff1]{SLAC National Accelerator Laboratory, 2575 Sand Hill Road, Menlo Park, CA 94025, U.S.A.}
\affil[aff2]{Currently at Brookhaven National Laboratory, Upton, NY 11973, U.S.A.}
%\affil[aff3]{You would list an author's second affiliation here.}
\corresp[cor1]{Corresponding author: blaj@slac.stanford.edu}
%\authornote[note1]{SLAC-PUB-17275}
%\authornote[note2]{This is an example of second authornote.}

\maketitle
\copyrightnotice

\begin{abstract}
ePix10K is a hybrid pixel detector developed at SLAC for demanding free-electron laser (FEL) applications, providing an ultrahigh dynamic range (\SI{245}{\electronvolt} to \SI{88}{\mega\electronvolt}) through gain auto-ranging. It has three gain modes (high, medium and low) and two auto-ranging modes (high-to-low and medium-to-low). The first ePix10K cameras are built around modules consisting of a sensor flip-chip bonded to 4 ASICs, resulting in \num{352x384} pixels of \SI{100}{\micro\metre}~x~\SI{100}{\micro\metre} each. We present results from extensive testing of three ePix10K cameras with FEL beams at LCLS, resulting in a measured noise floor of \SI{245}{\electronvolt} rms, or \SI{67}{\electron} equivalent noise charge (ENC), and a range of \num{11000} photons at \SI{8}{\kilo\electronvolt}. We demonstrate the linearity of the response in various gain combinations: fixed high, fixed medium, fixed low, auto-ranging high to low, and auto-ranging medium-to-low, while maintaining a low noise (well within the counting statistics), a very low cross-talk, perfect saturation response at fluxes up to \num{900} times the maximum range, and acquisition rates of up to \SI{480}{\hertz}. Finally, we present examples of high dynamic range x-ray imaging spanning more than 4 orders of magnitude dynamic range (from a single photon to \num{11000} photons/pixel/pulse at \SI{8}{\kilo\electronvolt}). Achieving this high performance with only one auto-ranging switch leads to relatively simple calibration and reconstruction procedures. The low noise levels allow usage with long integration times at non-FEL sources. ePix10K cameras leverage the advantages of hybrid pixel detectors with high production yield and good availability, minimize development complexity through sharing the hardware, software and DAQ development with all other versions of ePix cameras, while providing an upgrade path to \SI{5}{\kilo\hertz}, \SI{25}{\kilo\hertz} and \SI{100}{\kilo\hertz} in three steps over the next few years, matching the LCLS-II requirements.
\end{abstract}

% Head 1
\section{INTRODUCTION}

Free-electron lasers (FELs), with their unmatched combination of x-ray laser pulses  with energies (in the order of \si{\kilo\electronvolt}), time scales (in the order of femtoseconds), brilliance (\SIrange{E+12}{E+13}~\SI{8}{\kilo\electronvolt} photons) and coherence, similar to energy and time scales of atoms and molecules, enabled a new era of imaging the motion of atoms and molecules and the dynamics of chemical reactions. Currently there are four hard x-ray Free-Electron Lasers (FELs) operating worldwide: LCLS~(2009) \cite{emma2010first}, SACLA~(2011) \cite{pile2011x}, PAL-XFEL~(2016) and EuXFEL~(2017), with several more facilities planned to open in the near future (e.g., SwissFEL).

The short, high brilliance x-ray pulses at FELs are resulting in unique constraints on detectors: low noise, high range, good linearity, low cross-talk, high acquisition rates, detection of large numbers of photons entering the sensor within femtoseconds \cite{graafsma2009requirements}, and modularity for facilitating repairs \cite{blaj2016detector}. Typical FELs operate at \SIrange{60}{120}{\hertz}, while next generation FELs are planned to operate at much higher pulse rates: EuXFEL at \num{27000} pulses/s \cite{koch2013detector} and LCLS-2 gradually ramping up to \num{E6} pulses/second \cite{schoenlein2015new}, thus requiring a path towards high repetition rate detectors.

CSPAD detectors have been workhorse detectors at LCLS since the start in 2009, with tens of megapixels deployed in multiple cameras, and tens of petabytes of data collected \cite{hart2012cspad,herrmann2013cspad,blaj2014detector,blaj2015xray}. The ePix family represents a new generation of detectors developed at SLAC, greatly increasing all relevant parameters (noise, range, linearity, cross-talk) while offering an upgrade path towards \SI{5}{\kilo\hertz} and beyond \cite{blaj2015future}. The ePix100 detectors offer low noise (\SI{43}{\electron} equivalent noise charge, ENC, or \SI{155}{\electronvolt}~rms \cite{blaj2016xray}) and range of \num{100}~\SI{8}{\kilo\electronvolt} photons, with 12 cameras (0.5 megapixels each) deployed between 2015 and 2018 at LCLS \cite{carini2015the} and 4 cameras deployed at the EuXFEL. The ePix10K detectors are expanding the usable range while maintaining a low noise (single photon counting) through the use of different gain modes and auto-ranging \cite{caragiulo2014design}. Other gain auto-ranging, hybrid pixel detectors for FELs include Agipd (three gains, optimized for the particular pulse bunching structure of the EuXFEL beam \cite{henrich2011adaptive}) and Jungfrau (three gains \cite{mozzanica2014prototype}).

The ePix10K cameras are built around modules consisting of a sensor flip-chip bonded to 4 ASICs, resulting in \num{352x384} pixels of \SI{100x100}{\micro\metre} each (with a total area of \SI{35.2x38.4}{\milli\metre}). They are built on the ePix platform, sharing the hardware and software development with all other cameras built on the same platform \cite{dragone2014epix,nishimura2015design}.

Three ePix10K cameras have been extensively tested with FEL beams at LCLS in the various gain modes, resulting in a measured noise floor of \SI{67}{\electron} ENC (\SI{245}{\electronvolt}~rms) and a range of \num{11000}~\SI{8}{\kilo\electronvolt}~photons. We demonstrate the linearity of the response in various gain combinations: fixed high, fixed medium, fixed low, auto-ranging high to low, auto-ranging medium to low, while operating with very low cross-talk, and acquisition rates of up to \SI{480}{\hertz}.

\section{MATERIALS AND METHODS}

\subsection{ePix10K Camera}
\begin{figure}[t]
  \centerline {%
    \includegraphics[width =6.5in]{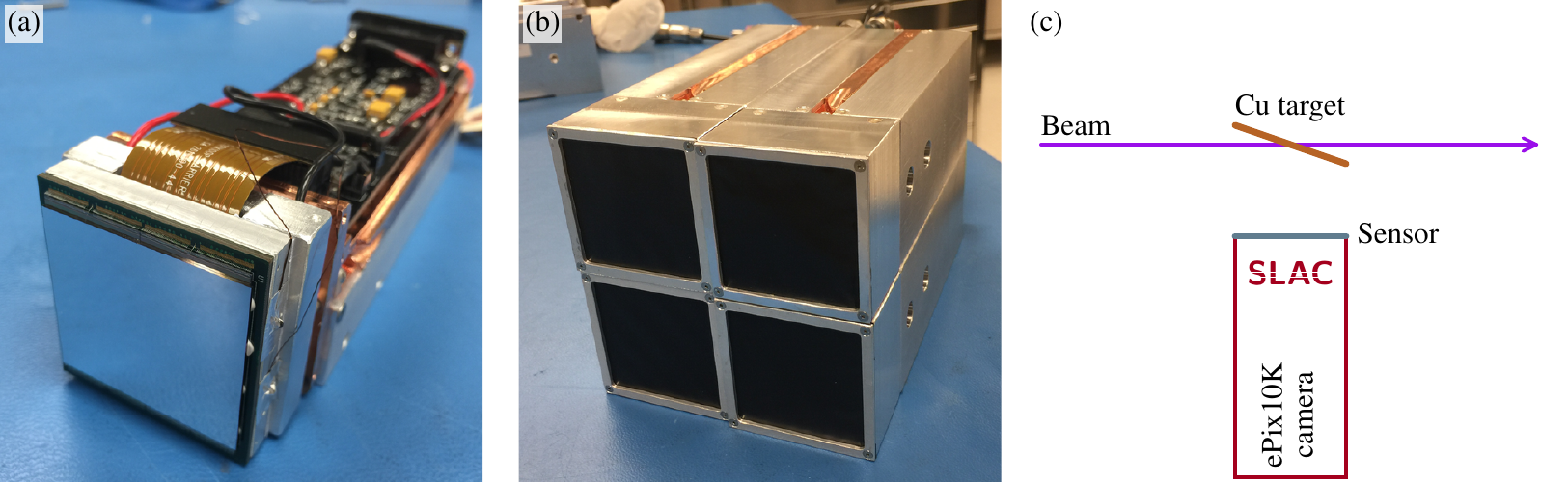}
  }
  \caption{(a)~Typical (vacuum compatible) ePix camera with covers removed, showing the large \SI{35.2x38.4}{\milli\metre} sensor with 4 read-out ASICs; (b)~the first four ePix10K cameras; (c)~top view schematic diagram of the experimental setup used for gain calibration, linearity testing, flat field and imaging tests; objects for imaging were inserted between the Cu target and the sensor.}
  \label{fig:a}
\end{figure}

ePix10K cameras are built around hybrid pixel detector modules obtained by flip-chip bonding 4 ePix10K readout ASICs \cite{caragiulo2014design, hart2014characterization} to a typical Si sensor (\SI{500}{\micro\metre} thick, n-type, resistivity \SI{10}{\kilo\ohm\centi\metre}; other options possible) \cite{blaj2017analytical}. For increased radiation hardness, the ASIC balconies are protected with high-Z strips \cite{tomada2012high}. The pixel size is \SI{100x100}{\micro\metre} and typical module size is \num{352x384}~pixels (\SI{35.2x38.4}{\milli\metre}). The ePix10K ASICs are designed on the standard ePix platform \cite{dragone2014epix} and the cameras are built on the ePix camera platform \cite{nishimura2015design} which is compatible with the different ePix readout chips. This results in fast development cycles and minimal integration effort with the LCLS data acquisition systems (DAQ), online monitoring (AMI) and offline analysis (psana). In Fig.~\ref{fig:a}~(a), a typical (vacuum-compatible) ePix camera is shown; (b)~depicts the first four ePix10K cameras.

The ePix10K readout ASICs enable high dynamic range imaging by implementing gain auto-ranging, i.e., in each frame, each pixel starts in a higher gain mode; if the pixel signal exceeds a configurable threshold, the pixel switches to a lower gain mode, increasing the range to match the incoming signal (similar to \cite{freytag2008kpix,henrich2011adaptive,mozzanica2014prototype}). This enables simultaneously low noise (high gain) acquisition of small signals and high range (low gain) acquisition of large signals, resulting in a dynamic range of \SI{255}{\electronvolt} to \SI{88}{\mega\electronvolt}. The ePix10K camera has 3 gain modes (high, medium, low), allowing 5 modes of operation (fixed high, fixed medium, fixed low, auto high to low, auto medium to low) which can be configured independently for each pixel. The different modes are summarized in Table~\ref{tab:a}. Each pixel value yields a 16 bit number, with the first 2 bits identifying the gain mode and the following 14 bits containing the digitized pixel output in the corresponding gain mode. Note the noise approaching \num{2}~analog-to-digital units (ADU) in medium and low gain, which is the expected quantization noise floor due to limited resolution of the analog-to-digital converter (ADC); a higher resolution ADC would enable further improvement of the signal to noise ratio.

\begin{table}[!t]
\caption{ePix10K operating modes and corresponding performance}
\label{tab:a}
\tabcolsep7pt
\begin{tabular}{ll|ccc|ccc|c}
\hline
\tch{2}{c}{b}{Gain mode} & \tch{3}{c}{b}{Gain} & \tch{3}{c}{b}{Noise} & \tch{1}{c}{b}{Range} \\

 & & \tch{1}{c}{b}{High} & \tch{1}{c}{b}{Medium} & \tch{1}{c}{b}{Low} & & & \\
 & & (ADU) & (ADU) & (ADU) & (ADU) & (eV) & (\si{\electron}) & (\SI{8}{\kilo\electronvolt} photons) \\
\hline
FH  & Fixed High            & 132 & -  & -    & 3.66 & 245         & 67 & 110                   \\
FM  & Fixed Medium          & -   & 43 & -    & 2.17 & 432         & 118 & 330                   \\
FL  & Fixed Low             & -   & -  & 1.32 & 2.00 & \num{12700} & \num{3480} & \num{11000}           \\
AHL & Auto High   to Low    & 133 & -  & 1.33 & 4.25 & 255         & 70 & \num{100}/\num{11000} \\
AML & Auto Medium to Low    & -   & 42 & 1.32 & 2.29 & 466         & 128 & \num{300}/\num{11000} \\
\hline
\end{tabular}
%\tablenote[t2n1]{Different gapless sensors can be used, e.g., 2x2, 2x1 ASICs per sensor}
%\tablenote[t2n2]{To convert to eV FWHM, multiply by 3.6*2.355}
%\tablenote[t2n3]{Cameras expected by LCLS-II first light}
\end{table}

The ePix10K operates currently at \SI{480}{\hertz} and offers an upgrade path in several steps to \SI{5}{\kilo\hertz}, \SI{25}{\kilo\hertz} and \SI{100}{\kilo\hertz}, matching the upcoming LCLS-II requirements.

\subsection{Experimental Setup}

The experiments were performed at the MFX beamline at LCLS \cite{boutet2016new}, which is optimized for molecular crystallography. The beam parameters were: photon beam energy \SI{9.5}{\kilo\electronvolt}, number of photons per pulse \num{E12}, pulse length \SI{42}{\femto\second}, repetition rate \SI{120}{\hertz}. The camera integrating time was set to \SI{100}{\micro\second}. For the characterization of the different gain modes, flat field calibration and imaging experiments, the cameras were mounted in the horizontal plane at an angle of \SI{90}{\degree} with the beam (to minimize Compton scattering), and a thin Cu target was placed in the beam, yielding Cu fluorescence (mostly K\textalpha{ }at \SI{8.05}{\kilo\electronvolt} and a small amount of K\textbeta{ }at \SI{8.90}{\kilo\electronvolt}); see diagram in Fig.~\ref{fig:a}~(c). The distance between the cameras and the interaction point was \SI{33}{\milli\metre} to achieve an average photon flux of \num{E5} photons/pixel/pulse. The average beam intensity was set (or scanned) by using an attenuator with 10 elements, resulting in an average beam intensity of \numrange{E-2}{E5}~photons/pixel/pulse. For each attenuator setting, the variations in FEL beam intensity allowed the collection of a range of different intensities. The beam intensity in each pulse was monitored and recorded with 6 beam monitoring diodes deployed at various locations around the cameras. For brevity, we will often quote a photon flux as a number of photons; all measurements in photons express the number of \SI{8}{\kilo\electronvolt} photons/pixel/FEL pulse. Throughout the paper, colors indicate the gain mode in which individual measurements were acquired: red represents high gain, green represents medium gain and blue represents low gain.

\section{RESULTS}

\subsection{Gain Calibration and Data Processing}
\begin{figure}[!h]
  \centerline {%
    \includegraphics[width =6.5in]{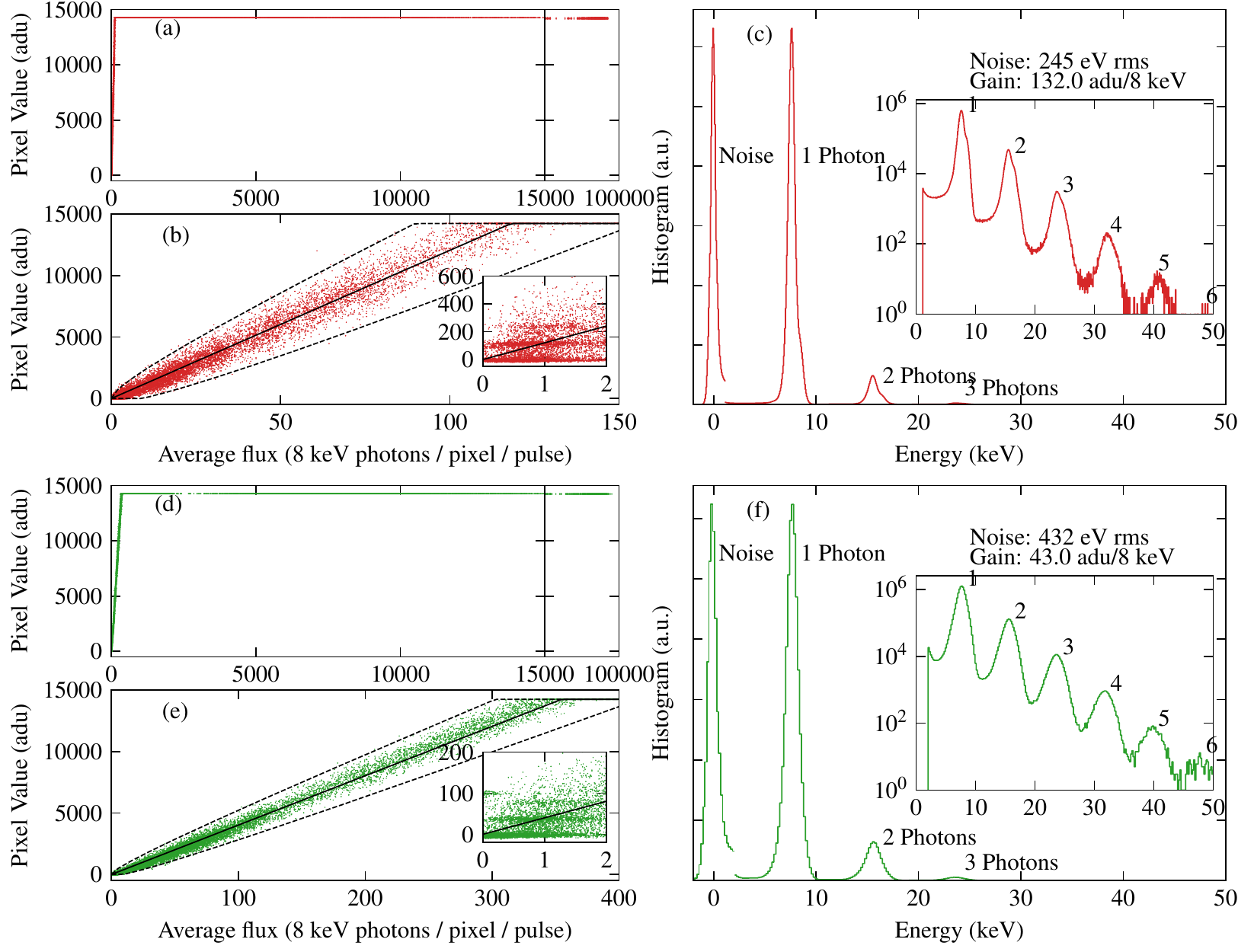}
  }
  \caption{Typical behavior of ePix10K camera in fixed high gain mode: (a) and (b) show the response of a single (randomly selected) pixel as a function of the average flux; each dot represents a single frame. (a) Shows the perfect saturation behavior in a range of \numrange{0}{E5} photons (yielding perfect saturation, despite exceeding the range by a factor \num{900}; note scale change between left and right panel at a flux of \num{15000} photons). (b) Displays a close-up of the region up to \numrange{0}{150} photons, demonstrating a range of \num{110} photons in high gain. The solid black line indicates the linear gain fit and the dashed lines indicate the corresponding Poisson counting statistics (using a stringent $\pm 3 \sigma$ noise criterion); note that almost all acquisitions are well within the counting statistics. The inset shows a close-up of the range of \numrange{0}{2}~photons. Note the vertical clustering of the dots corresponding to 0, 1, 2, etc. photon signals; also note the linearity over the whole range, with the fitting line passing through (0,0) and (1,1) photons while passing through the center of the acquisitions over the entire range. (c) Depicts the single pixel events spectrum obtained from all pixels in the camera (dark, common mode, gain correction, charge sharing rejection) and, for reference, the noise spectrum (dark and common mode corrected) scaled to match the 1 photon peak height; note the high signal to noise ratio and good peak separation. Inset shows the same on a logarithmic scale, demonstrating clear separation of up to 6 photons (only limited by statistics). Similarly, (d), (e), (f) depict results in fixed medium gain, with similar performance.}
  \label{fig:b}
\end{figure}

The gain calibration was performed with linear fits of individual pixel outputs as function of the average photon flux (measured with beam monitoring diodes), as indicated by the black line fits in Fig.~\ref{fig:b}~(b)~and~(e). In the high and medium gain modes, the gain was also calibrated by fitting the low occupancy histogram (noise peak and first 3 photon peaks) with a pixel charge sharing model, yielding the same results as full-range linear fits (note that ignoring the effects of charge sharing results in biased estimates of peak positions and incorrect gains in any pixel detector).

The data processing at low photon occupancy was performed by dark subtraction, common mode and gain correction, and histogramming. When applicable, single pixel events (i.e., photons depositing most of the charge in a single pixel) were selected from pixel values with signal $>3\sigma$ and 4 direct neighbors (up, down, left, right) within noise ($<3\sigma$) in individual frames and histogramming. At high occupancy, the common mode correction and single pixel event selection were not performed. All these operations were implemented with ultrafast Tensorflow algorithms (yielding same results as ``classical'', linear programming in 1-2 orders of magnitude shorter time) \cite{blaj2018ultrafast}.

A summary of all 5 modes of operation (fixed high, fixed medium, fixed low, auto-ranging high to low, auto-ranging medium to low), with the gains, noise and range results is presented in Table~\ref{tab:a}. Note that each pixel can be independently configured in a different mode of operation.

\subsection{Performance in Fixed Gain Modes}

In Fig.~\ref{fig:b}, the typical behavior in the fixed high and medium gain modes is depicted: (a) and (d) show the entire range of \numrange{0}{E5} photons, demonstrating the perfect saturation response, despite saturation over maximum range by factors of \num{900} and \num{300} for high and medium gain, respectively; (b) and (e) demonstrate the linear response of the high and medium gain modes over their entire range (well within the counting statistics bounds indicated by dashed lines), and the insets zoom in to the \numrange{0}{2} photons range where the vertical clustering indicates 0, 1, 2, etc. photons detected; note that the gain fit line passes through the centers of the clusters at the expected locations, demonstrating the consistent linearity from 1 photon to the full range. For brevity, and because singe photons are not visible in the low gain mode, the fixed low gain data is not shown.

Fig.~\ref{fig:b}~(c) and (f) detail the low light responses of the entire camera in fixed high and medium gain modes, demonstrating uniform responses, high signal to noise ratios and good photon separations up to 6 photons \cite{blaj2017optimal} (only limited by the size of the acquired dataset) for both fixed high and fixed medium gain modes. Note that the single pixel events spectrum is obtained from all pixels in the camera (after dark, common mode, gain correction, charge sharing rejection); the noise spectrum (dark and common mode corrected) is shown for reference, scaled to match the 1 photon peak height, resulting in an apparent discontinuity to the right of the noise peak. Note the high signal to noise ratio and good peak separation. The insets show the same data on a logarithmic scale.

The noise is (by definition) the rms width of the noise peak (e.g., \SI{255}{\electronvolt} in high gain); the width of the one photon peak is higher, due to the Fano noise (\SI{46}{\electronvolt} at \SI{5.9}{\kilo\electronvolt}, added quadratically) and the imperfect gain calibration (due to limited statistics); despite this, the K\textbeta{ }peak is just barely visible as the asymmetric shoulder to the right of the 1 photon peak in high gain mode (c).

The fixed gain modes offer slightly higher signal to noise ratios ($\sim$~\SI{4}{\percent} higher in fixed high gain compared to auto-ranging high to low and $\sim$~\SI{8}{\percent} higher in fixed medium gain compared to auto-ranging medium to low) and should be preferred when the range is known to be limited (under \num{110} photons in high gain and under \num{330} photon in medium gain). If higher (local or global) pixel intensities are expected, the auto-ranging gain modes are preferable.

\subsection{Performance in Auto-Ranging Gain Modes}

\begin{figure}[ht]
  \centerline {%
    \includegraphics[width =6.5in]{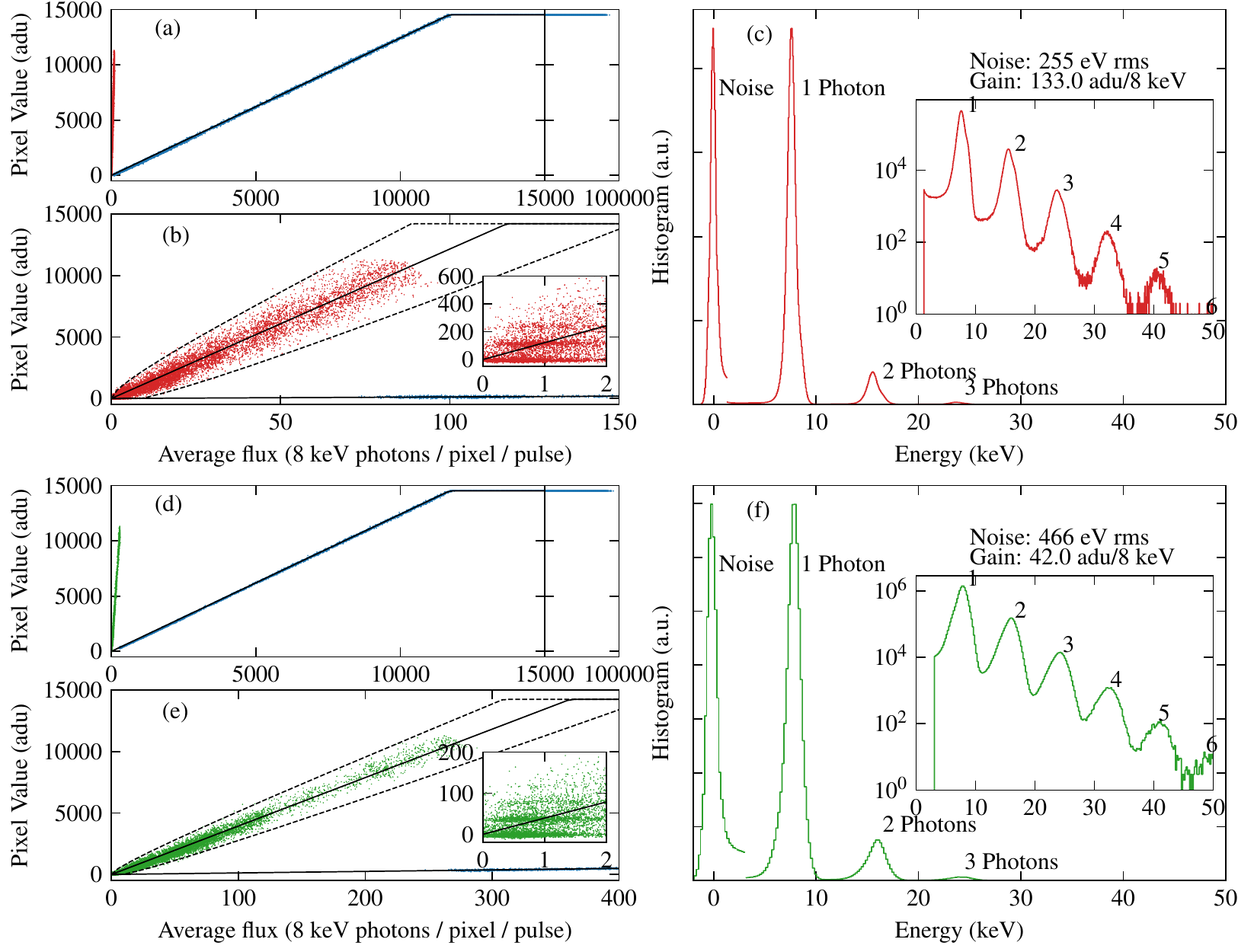}
  }
  \caption{Similar results to Fig.~\ref{fig:b} for auto ranging gain modes: (a), (b), (c) depict results for auto ranging high to low gain mode, while (d), (e) and (f) show results for auto ranging medium to low. Note the similar performance in high and medium gain, the additional low gain data (blue dots) and their linearity, and the auto ranging points around \num{90} photons in (b) and \num{270} photons around (e) for high and medium gain, respectively.}
  \label{fig:c}
\end{figure}

In auto-ranging modes, the high signal to noise ratio is maintained for the high and medium gain (with only modest increases in noise compared to the corresponding fixed gain modes, see previous paragraph; this slight increase in noise reflects a design trade-off between noise and maximum signal at pulsed sources). However, high intensities on individual pixels initiate a gain switch in those pixels, increasing their range to about \num{11000} photons. Figure~\ref{fig:c} shows the results, which are largely similar to the fixed gain modes shown in Fig.~\ref{fig:b}; note however the blue dots representing low gain measurements, and the auto-ranging of the gain, taking place at around \num{90} photons (b) and \num{270} photons (c). Also note the linear response in high gain, indicated by the matching of the blue dots and the corresponding black fit line. 

\begin{figure}[t]
  \centerline {%
    \includegraphics[width =6.5in]{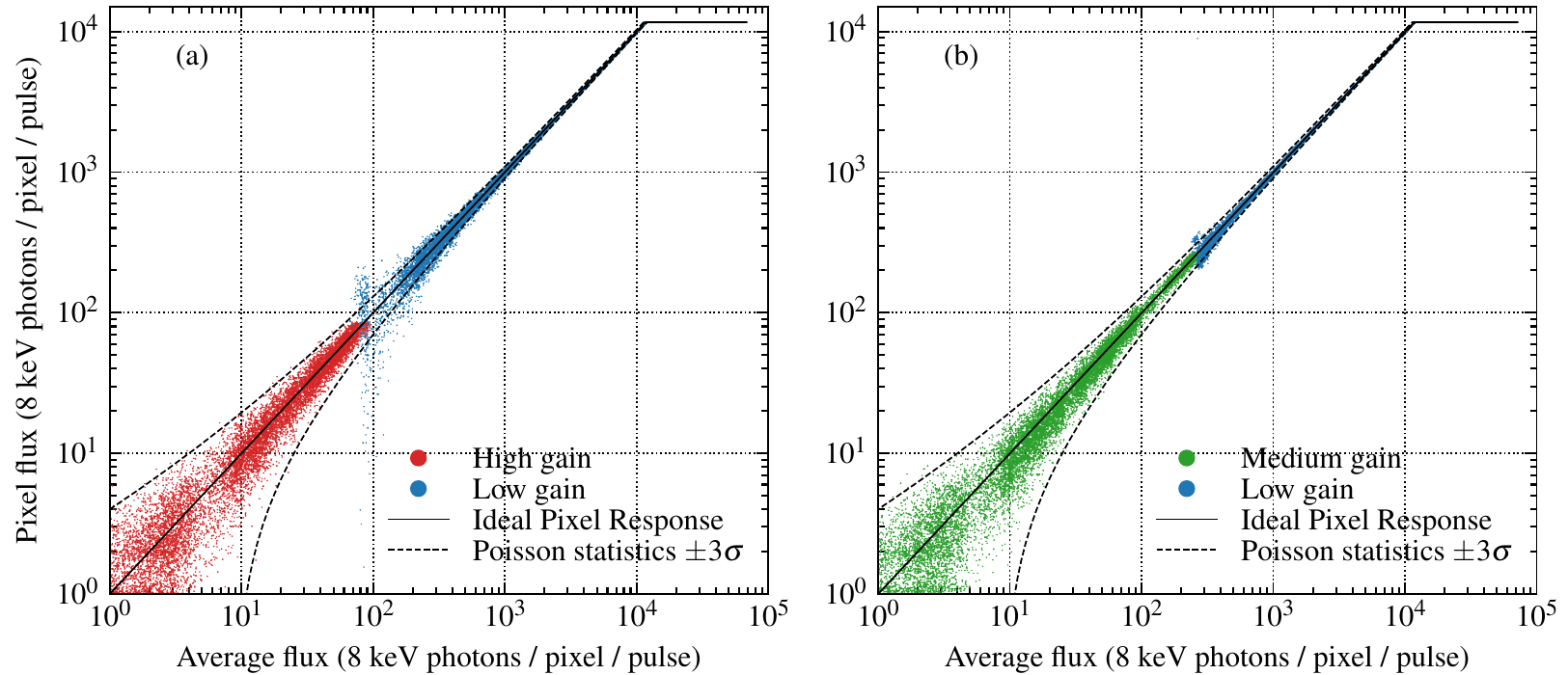}
  }
  \caption{ePix10K gain auto-ranging allows an ultra high dynamic range, with a noise of \SI{245}{\electronvolt}~rms (\SI{67}{\electron} ENC) and range of \num{11000}~\SI{8}{\kilo\electronvolt} photons; (a), (b) display the calibrated response of a single pixel as a function of average beam flux (measured with a separate beam monitoring diode) over \textgreater~4 orders of magnitude (log-log scale); red, green and blue dots represent high, medium and low gain measurements, respectively; the black line represents the ideal response; this demonstrates correct auto-ranging behavior, with good linearity and well within the photon counting limits (indicated by black dashed lines, using a stringent noise criterion of $\pm 3 \sigma$) over most of the range.}
  \label{fig:d}
\end{figure}

Because of the large dynamic ranges involved, the auto-ranging behavior between the different gain modes is difficult to show in a linear scale plot; in Fig.~\ref{fig:d} we show the auto gain ranging behavior on a log-log plot. (a) depicts the auto-ranging from high to low gain for one (randomly selected) pixel: red dots indicate high gain acquisitions, blue dots indicate low gain acquisitions, and the black line represents the ideal calibrated response. Dashed lines indicate the counting statistics (using a stringent criterion of $\pm 3 \sigma$); note that most dots are within the counting statistics. (b) shows the response of the same pixel in the auto-ranging medium to low gain mode, with similar performance. Gain auto-ranging in individual pixels results in small transients; in the most disadvantageous case (shown in Fig.~\ref{fig:d}), when the entire pixel matrix switches simultaneously, this results in additional noise right after the switching point; this effect will be minimized in the next generation of ePix for high rate applications, ePixHR. However, being able to choose between two different switching points (\num{90} and \num{270} photons shown here) allows the flexibility to choose the optimal measurement strategy for each experiment.

\subsection{X-ray Imaging Performance}

\begin{figure}[ht]
  \centerline {%
    \includegraphics[width =6.5in]{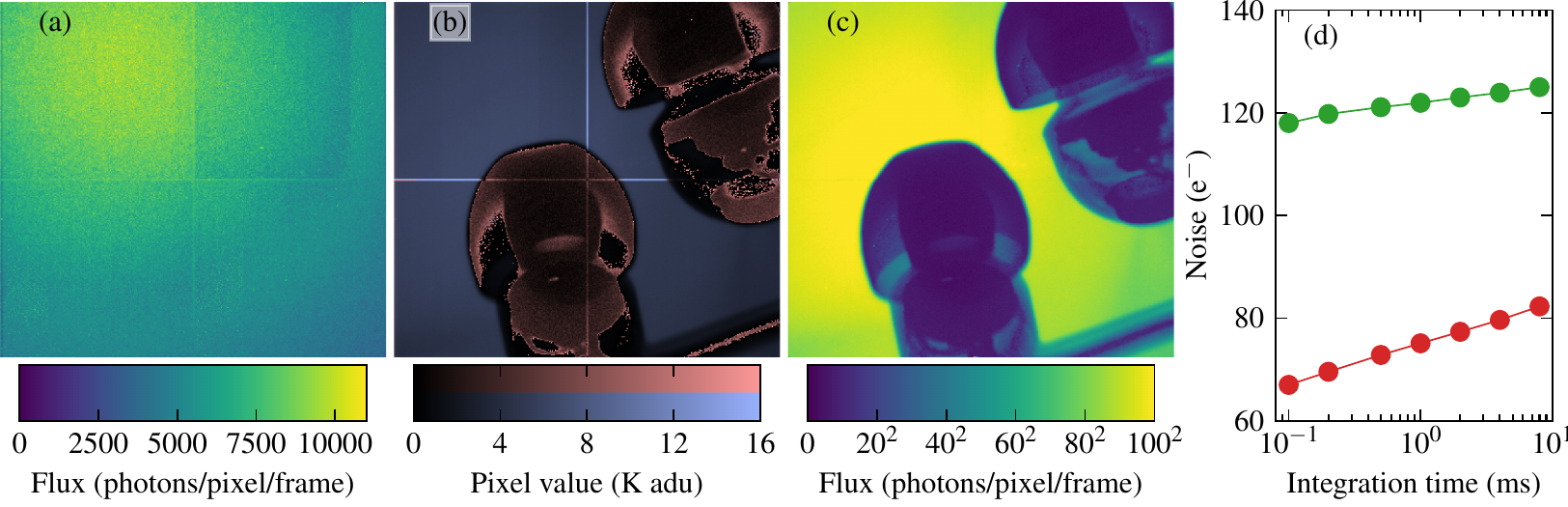}
  }
  \caption{ePix10K imaging and long integration noise results: (a) shows a single frame in auto-ranging medium to low gain and calibrated response; \SI{66}{\percent} of the pixels in are in low gain (top left corner) and the remaining pixels are in high gain, however, the two regions blend smoothly; (b) Displays the raw camera response (single frame, dark corrected) in auto-ranging high to low gain mode while imaging a pair of earbuds, with pixels behind dense image areas receiving a lower flux and reading out in high gain (red color), while pixels in the direct beam or behind thinner sample parts switched to low gain mode (blue color); (c) shows the corresponding calibrated response on a square root scale; note the high dynamic range (more than 4 orders of magnitude from single photons in the most dense sample areas to \num{11000} photons in the direct beam).  (d) Depicts the limited influence of integration time on noise performance: increasing the integration time from \SI{100}{\micro\second} to \SI{8}{\milli\second} only modestly increases the equivalent noise charge, from \SIrange{67}{82}{\electron} in fixed high gain (red dots) and from \SIrange{118}{125}{\electron} in fixed medium gain (green dots), enabling efficient detection in, e.g., synchrotron applications.}
  \label{fig:e}
\end{figure}

In Fig.~\ref{fig:e}, the functionality of the auto-ranging in high to low gain and medium to low gain modes is shown in several usage scenarios. Figure~\ref{fig:e}~(a) depicts a single, calibrated frame in auto medium to low gain of an illumination field with limited dynamic range (factor 2 in intensity between the brighter area in the top left quadrant and the darker area in the bottom right quadrant) and an average intensity close to the auto-ranging point (\num{270} photons), near the lowest point of signal to noise ratio in this operating mode; note however that the two regions are seamlessly blended.

A high dynamic range scene measured in auto-ranging high to low gain is depicted in Fig.~\ref{fig:e}~(b) and (c), covering more than 4 orders of magnitude, from single photons behind the most dense areas of the sample to the direct beam of \num{11000} photons. (b) Shows a raw image (single frame, pixel ADU values, dark corrected), with the red and blue colors indicating the high and low gain, respectively. The calibrated response (c) shows the seamless integration of the two gain areas, resulting in a high dynamic range image (image shown on a square root scale to facilitate inspection of both bright and dark areas simultaneously).

The limited influence of integration time on the noise performance is shown in Fig.~\ref{fig:e}~(d); increasing the integration time from \SI{100}{\micro\second} to \SI{8}{\milli\second} leads to relatively small increases of the ENC, from \SIrange{67}{82}{\electron} in fixed high gain (red dots) and from \SIrange{118}{125}{\electron} in fixed medium gain (green dots), enabling efficient detection in, e.g., synchrotron applications \cite{blaj2018hammerhead}; the low gain noise remains essentially unchanged as it is dominated by the ADC quantization noise (not shown in figure).

\section{CONCLUSIONS}

The ePix10K cameras have been developed at SLAC to meet demanding detector requirements for LCLS experiments (beyond the capabilities of existing detectors; see an overview in Table~\ref{tab:b}), while offering an upgrade path towards meeting LCLS-II requirements.

\begin{table}[!t]
\caption{Overview of SLAC detectors}
\label{tab:b}
\tabcolsep7pt
\begin{tabular}{lcccc}
\hline
 & \tch{1}{c}{b}{ePix10K} & \tch{1}{c}{b}{ePix100} & \tch{1}{c}{b}{ePixS} & \tch{1}{c}{b}{CSPAD}\\
 & & & & low / high gain \\
\hline
Summary & high dynamic range & low noise, deployed & spectroscopic & legacy, deployed\\
Mode of Operation & 2 gains, auto-ranging & 1 gain & 1 gain & 2 gains, fixed\\
Range (\SI{8}{\kilo\electronvolt} photons)& \num{11000} & \num{100} & \num{10}{}& \num{350} / \num{2700}  \\
Pixel size & \SI{100x100}{\micro\metre} & \SI{50x50}{\micro\metre} & \SI{500x500}{\micro\metre} & \SI{110x110}{\micro\metre} \\
Module size (pixels) & \num{352x384} & \num{704x768} & \num{20x20} & \num{370x388}\\
Noise (e$^{-}$ ENC) & \num{67}~e$^{-}$ & \num{43}~e$^{-}$ & \num{8}~e$^{-}$  & \num{300}~e$^{-}$ / \num{1000}~e$^{-}$\\
\hline
\end{tabular}
\end{table}

We demonstrated the performance of ePix10K cameras with LCLS beam, showing a low noise floor (\SI{67}{\electron} ENC or \SI{245}{\electronvolt} rms), performance and linearity in the 5 operation modes (fixed high, fixed medium, fixed low, auto-ranging high to low, and auto-ranging medium-to-low, configurable pixel by pixel), correct auto-ranging behavior and tolerance to photon fluxes exceeding the maximum range by up to \num{900} times.

Achieving this high performance with only one auto-ranging switch (i.e., two gains) leads to relatively simple calibration and reconstruction procedures (compared to, e.g., \cite{mezza2016new, redford2018first}). The low noise levels allow usage with long integration times at non-FEL sources. Finally, we presented examples of high dynamic range x-ray imaging spanning more than 4 orders of magnitude dynamic range (from single photons to \num{11000}~photons/pixel/pulse).

ePix10K cameras are relatively inexpensive, leveraging the advantages of hybrid pixel detectors with high production yield and good availability. They are modular, easy to scale and integrate \cite{blaj2016detector} and offer an upgrade path in several steps to \SI{5}{\kilo\hertz}, \SI{25}{\kilo\hertz} and \SI{100}{\kilo\hertz}.

% Sections that will go in second font

% Acknowledgement
\section{ACKNOWLEDGMENTS}
Use of the Linac Coherent Light Source (LCLS), SLAC National Accelerator Laboratory, is supported by the U.S. Department of Energy, Office of Science, Office of Basic Energy Sciences under Contract No. DE-AC02-76SF00515. Publication number SLAC-PUB-17275.

% References

%\bibliographystyle{aipnum-cp}%
%\bibliography{main}%
%Control: year  (1) truncated

\end{document}